\documentclass[11pt,twoside]{article}
\usepackage{asp2010}

\resetcounters

\begin{document}

\title{Observations and modeling of the massive young star AFGL~4176:
  From large scales to small} 

\author{Paul~A.~Boley,$^1$ Roy~van~Boekel,$^1$ Hendrik~Linz,$^1$
  Jeroen~Bouwman,$^1$ Andrey~M.~Sobolev$^2$ and Thomas~Henning$^1$
  \affil{$^1$Max Planck Institute for Astronomy, Heidelberg,
    Germany\\ $^2$Ural Federal University, Astronomical Observatory,
    Ekaterinburg, Russia}}

\begin{abstract}
We present spatially-resolved mid-infrared interferometric
observations of the massive young stellar object AFGL~4176, together
with literature and survey data.  We model these observations using a
simple, spherically-symmetric radiative transfer model, and find that
the observational data are consistent with a highly luminous star
surrounded by a thick envelope.
\end{abstract}

\section{Introduction}

Massive stars in the act of formation are a rare phenomenon in our
galaxy.  Characterization of these stars, in the form of massive young
stellar objects (MYSOs), is required to constrain formation and
evolution scenarios.  However, any successful observational strategy
must account for the inherently large distances (several kpc) and
extremely high extinctions ($A_V \sim 100$~mag) involved.  Finally,
interpretation of observations almost universally requires the
construction or adaption of an underlying model.

In an effort to understand such objects better, many authors attempt
to reproduce the spectral energy distribution (SED) without accounting
for the spatial distribution of emission, as spatially-resolved
observations are often not available.  However, such an approach is
highly degenerate \citep[e.g.][]{Menshchikov97}.  Here, we combine
extensive multi-wavelength observations with numerical radiative
transfer modeling for a single MYSO candidate, AFGL~4176.  Our
approach encompasses both spectral information in the form of the
spectral energy distribution (SED), and spatial information in the
form of mid-infrared interferometric visibilities and resolved imaging
at far-infrared and sub-millimeter wavelengths.

AFGL~4176 (also known as IRAS~13395-6153, G308.9+0.1) was identified
by \citet{Henning84} as a candidate MYSO based on the similarity of
the far-infrared spectrum to that of the Becklin-Neugebauer (BN)
object in Orion.  However, this object differs from other BN-type
objects by virtue of the extremely high luminosity of the central
source, with previous works based on SED-fitting suggesting the
luminosity is on the order of $10^5$~L$_{\sun}$
\citep{Guertler91,Grave09}.  From 1.2~mm continuum measurements,
\citet{Beltran06} estimated a gas + dust mass of 1120~M$_{\sun}$.
Regarding the evolutionary status of AFGL~4176, a search for an
outflow by \citet{DeBuizer09} was inconclusive, although an
ultra-compact \ion{H}{II} region is present
\citep[e.g.][]{Phillips98}.  To date, no compelling evidence for the
presence of an accretion disk has been presented.

The distance to AFGL~4176 remains uncertain, as the combination of
$V_\mathrm{LSR}$ for the source and galactic coordinates are forbidden
in most galaxy rotation curves.  However, for the purpose of this
work, we adopt a distance of 3200~pc (A. V. Loktin, private
communication).

\begin{figure}[!t]
\plotone{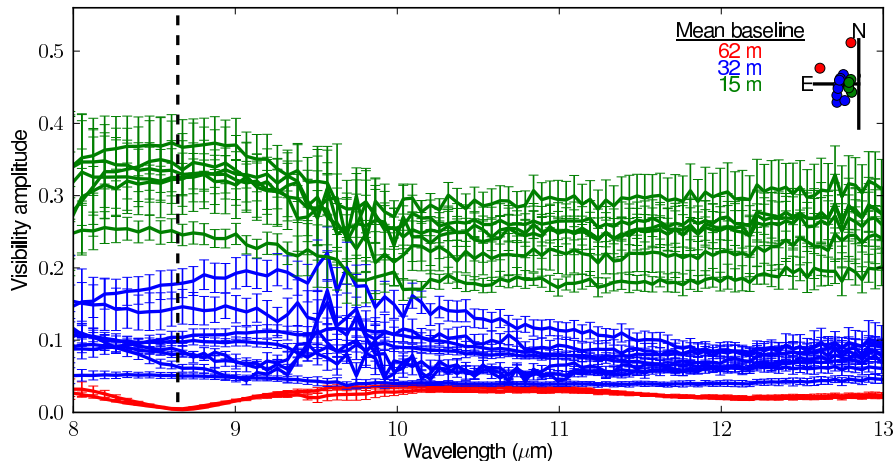}
\caption{Visibility amplitudes from MIDI.  Measurements are colored by
  projected baseline length, and the locations in UV space are shown
  in the top right.  The dashed line shows the location of the zero
  crossing observed at long baselines (red).}
\label{fig_visamp}
\end{figure}

\begin{figure}[!t]
\plotone{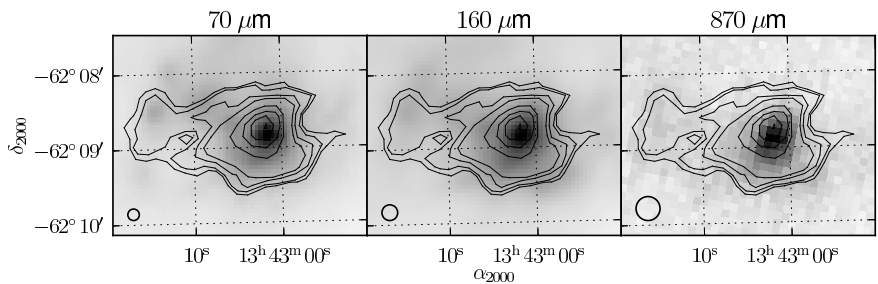}
\caption{Far-IR and sub-mm images of the region around AFGL~4176 from
  the Hi-Gal and ATLASGAL surveys.  The contours show the 1.2~mm
  emission, and the circles show the approximate FWHM beam sizes.  See
  text for details.}
\label{fig_image}
\end{figure}

\section{Observations}

We observed AFGL~4176 as part of a guaranteed time for observations
program using the two-telescope mid-infrared interferometer MIDI on
the Very Large Telescope of the European Southern Observatory.  We
present 19 spectrally-resolved visibility measurements in the $N$ band
($8 - 13$~$\mu$m), obtained in the period from 2005 to 2007, which we
show in Fig.~\ref{fig_visamp}.  The projected baselines span from 13
to 62~m, which corresponds to spatial scales $\lambda/2B$ of about 50
to 250 AU.

In Fig.~\ref{fig_image}, we present images of the region at 70 and
160~$\mu$m, obtained with the Herschel space telescope as part of the
Hi-Gal program \citep{Molinari10}, and at 870~$\mu$m, obtained with
the Atacama Pathfinder EXperiment (APEX) telescope as part of the
ATLASGAL survey \citep{Schuller09}.  The source is resolved at these
three wavelengths, though shows little structure.  For comparison, we
have overlaid contours of the 1.2~mm emission measured by
\citet{Beltran06}, obtained with the Swedish-ESO Sub-millimeter
Telescope (SEST).

\begin{figure}[!t]
\plotone{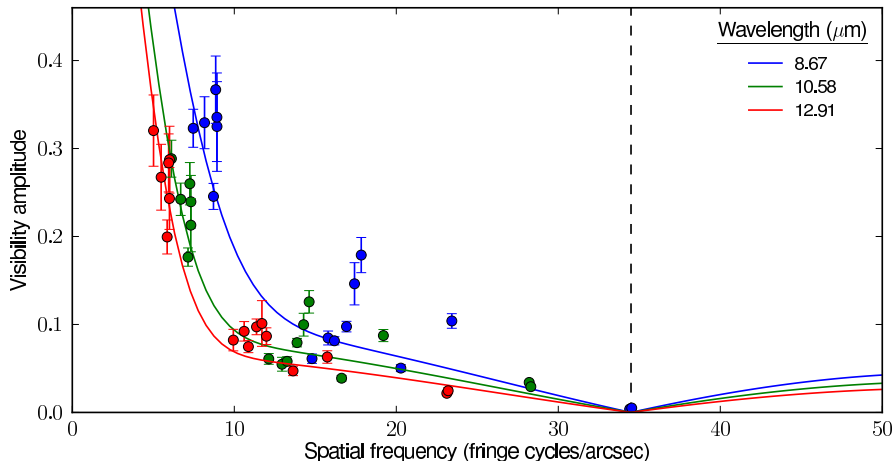}
\caption{Geometric fits to the visibility amplitudes at three
  different wavelengths.  The dashed line indicates the spatial
  frequency where the zero crossing is observed.}
\label{fig_visfit}
\end{figure}

\section{Geometric models of $N$-band visibilities}

On two equivalent projected baselines (62~m; red lines in
Fig.~\ref{fig_visamp}) at position angles differing by 63~deg, we
observe a zero crossing in the visibility amplitudes accompanied by a
phase flip in the differential phase (not presented).  This
phenomenon, marked with a dashed line in Fig.~\ref{fig_visamp}, not
only implies the presence of some sort of hard ``edge'' in the source
intensity distribution, but also suggests that the source is
circularly symmetric on the smallest scales probed by our
observations.

We attempt to model this behavior with a simple, circularly-symmetric
geometric model, consisting of a Gaussian and a thin ring.  Using the
spatial frequency of the zero-crossing as a constraint for the size of
the ring, we find that the observed visibilities can be roughly
reproduced by a 100~mas ($\sim$300~AU) extended component combined
with a compact, 10~mas ($\sim$30~AU) ring-like component.  We show the
results of this geometric fit in Fig.~\ref{fig_visfit}.

\section{Radiative transfer modeling}

We model the observational data (SED, MIDI visibilities, far-infrared
and sub-mm radial intensity profiles) from near-IR through mm
wavelengths using a one-dimensional radiative transfer code.  We
employ a simple dust model consisting of opacities from
\citet{Ossenkopf94} for the cold dust ($T_\mathrm{dust} < 130$~K).
For the remaining dust ($T_\mathrm{dust} \geq 130$~K), we use a 60\%/40\%
mix of astronomical silicates and graphite with an MRN size
distribution, using opacities from \citet{Laor93}.

We find that a piecewise power law for the density distribution is
able to adequately reproduce the SED, mid-IR visibilities and
far-IR/sub-mm spatial intensity profiles.  Around a 15\,000~K black
body central source with a luminosity of $1.4 \times 10^5$~L$_{\sun}$,
we have an envelope with a dust mass of 5.2~M$_{\sun}$.  The dust
density distribution $\rho(r)$ consists of a hot inner shell with
enhanced density, spanning from 40 to 65~AU, surrounded by an extended
envelope out to 125\,000~AU, with a change in the density falloff at
70\,000~AU:
\begin{displaymath}
\rho(r) = \left\{ \begin{array}{ll}
  \left(2.6 \times 10^{-18} \textrm{~g cm}^{-3}\right) \left(\frac{40~\textrm{\scriptsize AU}}{r}\right)^{1.8} & 40~\textrm{AU} < r \leq 65~\textrm{AU}\\
  \left(3.9 \times 10^{-22} \textrm{~g cm}^{-3} \right) \left(\frac{70\,000~\textrm{\scriptsize AU}}{r}\right)^{1.0} & 65~\textrm{AU} < r \leq 70\,000~\textrm{AU} \\
  \left(3.9 \times 10^{-22} \textrm{~g cm}^{-3} \right) \left(\frac{70\,000~\textrm{\scriptsize AU}}{r}\right)^{0.5} & 70\,000~\textrm{AU} < r \leq 125\,000~\textrm{AU}
  \end{array} \right.
\end{displaymath}

\begin{figure}[!t]
\plotone{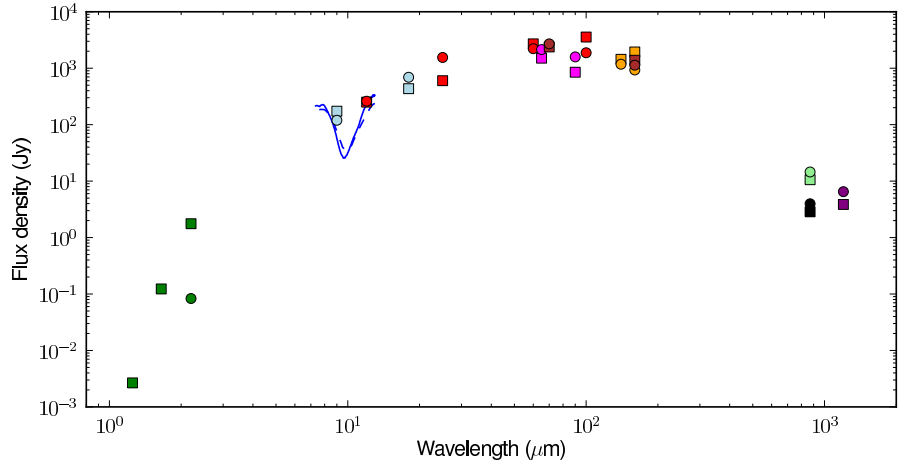}
\caption{Comparison of the radiative transfer model fluxes (circles
  and dashed line) to observations (squares and solid line).  Aperture
  and beam sizes for the measurements have been taken into account.
  In nearly all cases, error bars are smaller than the marker size.}
\label{fig_sed}
\end{figure}

\begin{figure}[!t]
\plottwo{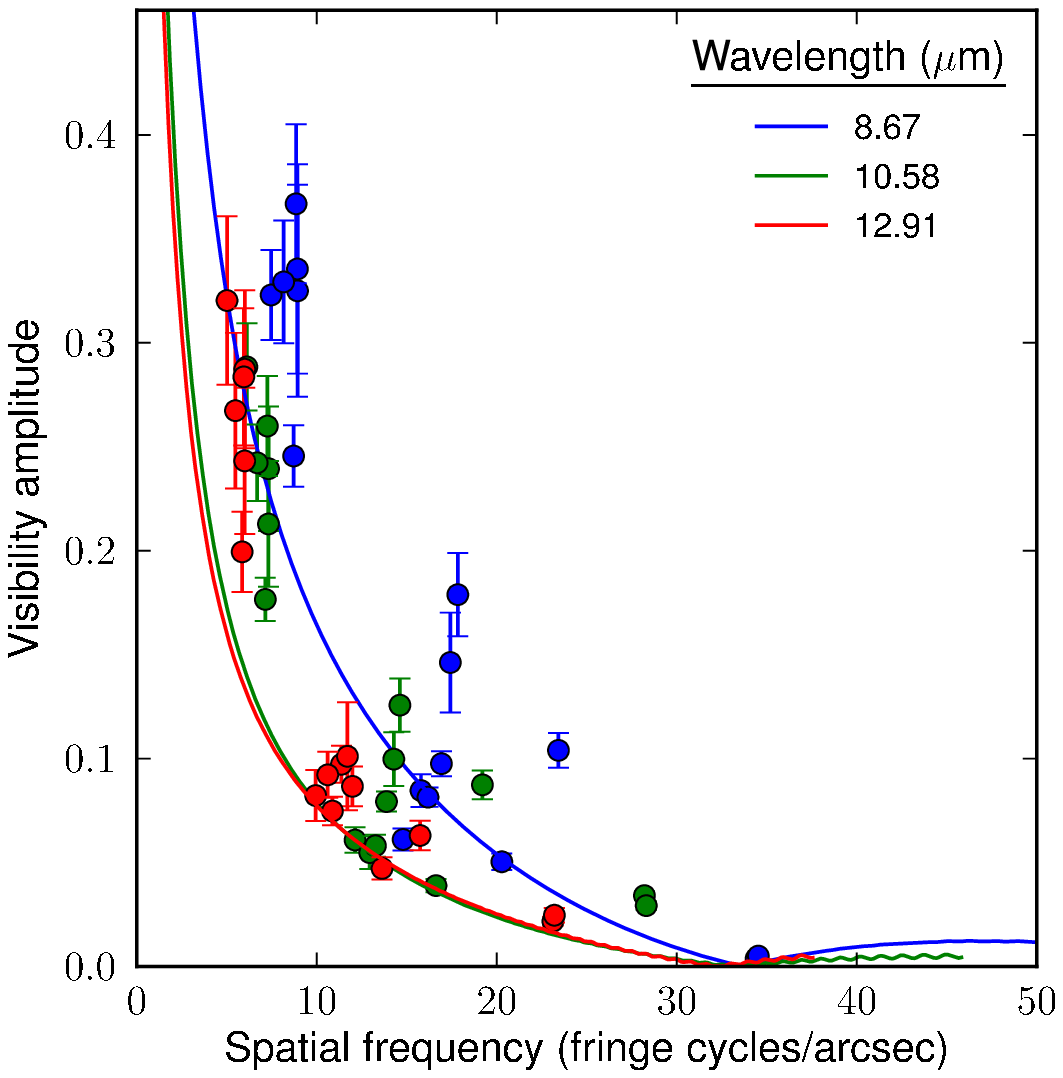}{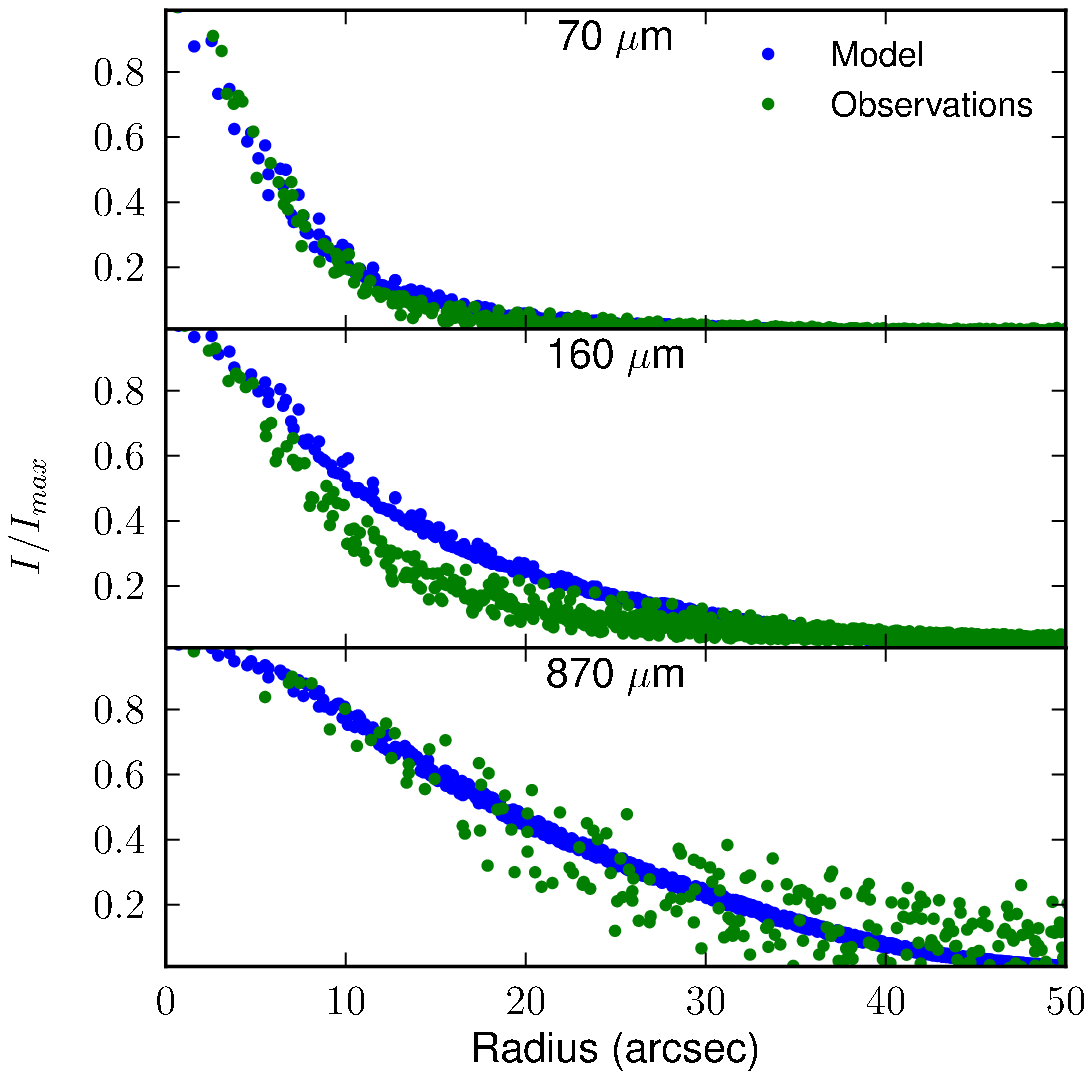}
\caption{\emph{Left:} Radiative transfer model visibilities (lines)
  and observed MIDI visibilities (points) at three selected
  wavelengths. \emph{Right:} Spatial profiles from the radiative
  transfer model (blue points) compared with observations from
  Herschel and APEX (green points).}
\label{fig_spatial}
\end{figure}

Our model fit to the SED is shown in Fig.~\ref{fig_sed}, where the
effects of aperture and beam sizes on the measured flux levels have
been accounted for.  The resulting visibilities and the PSF-convolved
radial intensity profiles at 70, 160 and 870~$\mu$m are shown in
Fig.~\ref{fig_spatial}.  On the whole, the SED and radial profiles can
be reproduced well by a spherical model.  Near-infrared fluxes are
underestimated (similar to the models of \citet{Guertler91} and
\citet{Siebenmorgen93}), and this may be a fundamental limitation of a
spherical model.  Despite the simplicity of the density structure
used, the model visibility amplitudes match observations rather well.
However, a better fit may be possible with a more complex density
structure at scales of $\sim$100~AU.

\section{Summary and conclusion}

We have simultaneously fit the SED and spatial structure at both large
($10^5$~AU) and small ($10^1$~AU) scales of the massive young stellar
object AFGL~4176 using a spherical radiative transfer model and an
appropriate dust model.  The simple density structure used and
spherical geometry are enough to reproduce many spectral and spatial
details in the observations of AFGL~4176.

In the present case, there are no clear signs of significant
deviations from spherical symmetry.  However, whether this is the
result of a face-on orientation or indicative of a true spheroidal
nature remains untested.  In particular, the physical nature of the
density discontinuity at $\sim65$~AU could be quite different than
presented here (e.g. cool material shaded by the rim of a disk).

Finally, we emphasize that \emph{only} when spatially-resolved
observations (at both large and small scales) are combined with SED
data may we derive the true physical structure of such objects,
whereas modeling of the structure based on SED data alone is
degenerate.  However, the uniqueness (or ubiquity) of a possible model
is nonetheless a strong function of spatial coverage.

\acknowledgements We thank the organizers of the MRO interferometry
workshop for providing an excellent forum to discuss recent results
and advancements in the field of optical interferometry.

\bibliography{boley}

\begin{thebibliography}{}
\expandafter\ifx\csname natexlab\endcsname\relax\def\natexlab#1{#1}\fi
\expandafter\ifx\csname url\endcsname\relax
  \def\url#1{\texttt{#1}}\fi
\expandafter\ifx\csname urlprefix\endcsname\relax\def\urlprefix{URL }\fi
\providecommand{\eprint}[2][]{\url{#2}}

\bibitem[{{Beltr{\'a}n} et~al.(2006){Beltr{\'a}n}, {Brand}, {Cesaroni},
  {Fontani}, {Pezzuto}, {Testi}, \& {Molinari}}]{Beltran06}
{Beltr{\'a}n}, M.~T., {Brand}, J., {Cesaroni}, R., {Fontani}, F., {Pezzuto},
  S., {Testi}, L., \& {Molinari}, S. 2006, \aap, 447, 221.
  \eprint{arXiv:astro-ph/0510422}

\bibitem[{{De Buizer} et~al.(2009){De Buizer}, {Redman}, {Longmore}, {Caswell},
  \& {Feldman}}]{DeBuizer09}
{De Buizer}, J.~M., {Redman}, R.~O., {Longmore}, S.~N., {Caswell}, J., \&
  {Feldman}, P.~A. 2009, \aap, 493, 127. \eprint{0810.4951}

\bibitem[{{Grave} \& {Kumar}(2009)}]{Grave09}
{Grave}, J.~M.~C., \& {Kumar}, M.~S.~N. 2009, \aap, 498, 147.
  \eprint{0901.2053}

\bibitem[{{Guertler} et~al.(1991){Guertler}, {Henning}, {Kruegel}, \&
  {Chini}}]{Guertler91}
{Guertler}, J., {Henning}, T., {Kruegel}, E., \& {Chini}, R. 1991, \aap, 252,
  801

\bibitem[{{Henning} et~al.(1984){Henning}, {Friedemann}, {Guertler}, \&
  {Dorschner}}]{Henning84}
{Henning}, T., {Friedemann}, C., {Guertler}, J., \& {Dorschner}, J. 1984,
  Astronomische Nachrichten, 305, 67

\bibitem[{{Laor} \& {Draine}(1993)}]{Laor93}
{Laor}, A., \& {Draine}, B.~T. 1993, \apj, 402, 441

\bibitem[{{Men'shchikov} \& {Henning}(1997)}]{Menshchikov97}
{Men'shchikov}, A.~B., \& {Henning}, T. 1997, \aap, 318, 879

\bibitem[{{Molinari} et~al.(2010){Molinari}, {Swinyard}, {Bally}, {Barlow},
  {Bernard}, {Martin}, {Moore}, {Noriega-Crespo}, {Plume}, {Testi}, {Zavagno},
  {Abergel}, {Ali}, {Andr{\'e}}, {Baluteau}, {Benedettini}, {Bern{\'e}},
  {Billot}, {Blommaert}, {Bontemps}, {Boulanger}, {Brand}, {Brunt}, {Burton},
  {Campeggio}, {Carey}, {Caselli}, {Cesaroni}, {Cernicharo}, {Chakrabarti},
  {Chrysostomou}, {Codella}, {Cohen}, {Compiegne}, {Davis}, {de Bernardis}, {de
  Gasperis}, {Di Francesco}, {di Giorgio}, {Elia}, {Faustini}, {Fischera},
  {Fukui}, {Fuller}, {Ganga}, {Garcia-Lario}, {Giard}, {Giardino}, {Glenn},
  {Goldsmith}, {Griffin}, {Hoare}, {Huang}, {Jiang}, {Joblin}, {Joncas},
  {Juvela}, {Kirk}, {Lagache}, {Li}, {Lim}, {Lord}, {Lucas}, {Maiolo},
  {Marengo}, {Marshall}, {Masi}, {Massi}, {Matsuura}, {Meny}, {Minier},
  {Miville-Desch{\^e}nes}, {Montier}, {Motte}, {M{\"u}ller}, {Natoli}, {Neves},
  {Olmi}, {Paladini}, {Paradis}, {Pestalozzi}, {Pezzuto}, {Piacentini},
  {Pomar{\`e}s}, {Popescu}, {Reach}, {Richer}, {Ristorcelli}, {Roy}, {Royer},
  {Russeil}, {Saraceno}, {Sauvage}, {Schilke}, {Schneider-Bontemps},
  {Schuller}, {Schultz}, {Shepherd}, {Sibthorpe}, {Smith}, {Smith},
  {Spinoglio}, {Stamatellos}, {Strafella}, {Stringfellow}, {Sturm}, {Taylor},
  {Thompson}, {Tuffs}, {Umana}, {Valenziano}, {Vavrek}, {Viti}, {Waelkens},
  {Ward-Thompson}, {White}, {Wyrowski}, {Yorke}, \& {Zhang}}]{Molinari10}
{Molinari}, S., {Swinyard}, B., {Bally}, J., {Barlow}, M., {Bernard}, J.-P.,
  {Martin}, P., {Moore}, T., {Noriega-Crespo}, A., {Plume}, R., {Testi}, L.,
  {Zavagno}, A., {Abergel}, A., {Ali}, B., {Andr{\'e}}, P., {Baluteau}, J.-P.,
  {Benedettini}, M., {Bern{\'e}}, O., {Billot}, N.~P., {Blommaert}, J.,
  {Bontemps}, S., {Boulanger}, F., {Brand}, J., {Brunt}, C., {Burton}, M.,
  {Campeggio}, L., {Carey}, S., {Caselli}, P., {Cesaroni}, R., {Cernicharo},
  J., {Chakrabarti}, S., {Chrysostomou}, A., {Codella}, C., {Cohen}, M.,
  {Compiegne}, M., {Davis}, C.~J., {de Bernardis}, P., {de Gasperis}, G., {Di
  Francesco}, J., {di Giorgio}, A.~M., {Elia}, D., {Faustini}, F., {Fischera},
  J.~F., {Fukui}, Y., {Fuller}, G.~A., {Ganga}, K., {Garcia-Lario}, P.,
  {Giard}, M., {Giardino}, G., {Glenn}, J.~., {Goldsmith}, P., {Griffin}, M.,
  {Hoare}, M., {Huang}, M., {Jiang}, B., {Joblin}, C., {Joncas}, G., {Juvela},
  M., {Kirk}, J., {Lagache}, G., {Li}, J.~Z., {Lim}, T.~L., {Lord}, S.~D.,
  {Lucas}, P.~W., {Maiolo}, B., {Marengo}, M., {Marshall}, D., {Masi}, S.,
  {Massi}, F., {Matsuura}, M., {Meny}, C., {Minier}, V.,
  {Miville-Desch{\^e}nes}, M.-A., {Montier}, L., {Motte}, F., {M{\"u}ller},
  T.~G., {Natoli}, P., {Neves}, J., {Olmi}, L., {Paladini}, R., {Paradis}, D.,
  {Pestalozzi}, M., {Pezzuto}, S., {Piacentini}, F., {Pomar{\`e}s}, M.,
  {Popescu}, C.~C., {Reach}, W.~T., {Richer}, J., {Ristorcelli}, I., {Roy}, A.,
  {Royer}, P., {Russeil}, D., {Saraceno}, P., {Sauvage}, M., {Schilke}, P.,
  {Schneider-Bontemps}, N., {Schuller}, F., {Schultz}, B., {Shepherd}, D.~S.,
  {Sibthorpe}, B., {Smith}, H.~A., {Smith}, M.~D., {Spinoglio}, L.,
  {Stamatellos}, D., {Strafella}, F., {Stringfellow}, G., {Sturm}, E.,
  {Taylor}, R., {Thompson}, M.~A., {Tuffs}, R.~J., {Umana}, G., {Valenziano},
  L., {Vavrek}, R., {Viti}, S., {Waelkens}, C., {Ward-Thompson}, D., {White},
  G., {Wyrowski}, F., {Yorke}, H.~W., \& {Zhang}, Q. 2010, \pasp, 122, 314.
  \eprint{1001.2106}

\bibitem[{{Ossenkopf} \& {Henning}(1994)}]{Ossenkopf94}
{Ossenkopf}, V., \& {Henning}, T. 1994, \aap, 291, 943

\bibitem[{{Phillips} et~al.(1998){Phillips}, {Norris}, {Ellingsen}, \&
  {McCulloch}}]{Phillips98}
{Phillips}, C.~J., {Norris}, R.~P., {Ellingsen}, S.~P., \& {McCulloch}, P.~M.
  1998, \mnras, 300, 1131

\bibitem[{{Schuller} et~al.(2009){Schuller}, {Menten}, {Contreras}, {Wyrowski},
  {Schilke}, {Bronfman}, {Henning}, {Walmsley}, {Beuther}, {Bontemps},
  {Cesaroni}, {Deharveng}, {Garay}, {Herpin}, {Lefloch}, {Linz}, {Mardones},
  {Minier}, {Molinari}, {Motte}, {Nyman}, {Reveret}, {Risacher}, {Russeil},
  {Schneider}, {Testi}, {Troost}, {Vasyunina}, {Wienen}, {Zavagno}, {Kovacs},
  {Kreysa}, {Siringo}, \& {Wei{\ss}}}]{Schuller09}
{Schuller}, F., {Menten}, K.~M., {Contreras}, Y., {Wyrowski}, F., {Schilke},
  P., {Bronfman}, L., {Henning}, T., {Walmsley}, C.~M., {Beuther}, H.,
  {Bontemps}, S., {Cesaroni}, R., {Deharveng}, L., {Garay}, G., {Herpin}, F.,
  {Lefloch}, B., {Linz}, H., {Mardones}, D., {Minier}, V., {Molinari}, S.,
  {Motte}, F., {Nyman}, L., {Reveret}, V., {Risacher}, C., {Russeil}, D.,
  {Schneider}, N., {Testi}, L., {Troost}, T., {Vasyunina}, T., {Wienen}, M.,
  {Zavagno}, A., {Kovacs}, A., {Kreysa}, E., {Siringo}, G., \& {Wei{\ss}}, A.
  2009, \aap, 504, 415. \eprint{0903.1369}

\bibitem[{{Siebenmorgen}(1993)}]{Siebenmorgen93}
{Siebenmorgen}, R. 1993, \apj, 408, 218

\end{thebibliography}

\end{document}